\documentclass[prb,superscriptaddress,showpacs,amsmath,amssymb,twocolumn]{revtex4-2}

\usepackage{graphicx}
\usepackage{dcolumn}
\usepackage{bm}
\usepackage{epstopdf}
\usepackage{upgreek}
\usepackage{amsmath,amssymb,amsthm,commath,esint,tikz-cd,tikz, mathtools}
\usepackage{array}

\numberwithin{definition}{section}

\numberwithin{proposition}{section}

\numberwithin{theorem}{section}

\numberwithin{lemma}{section}

\numberwithin{corollary}{section}
\theoremstyle{remark}

\numberwithin{example}{section}

\usepackage{comment}

\begin{document}

\title{Phonon-mediated superconductivity in ternary silicides $X_4$CoSi ($X=$ Nb, Ta)}

\author{Jonas Bekaert}
\email{jonas.bekaert@uantwerpen.be}
\affiliation{%
 Department of Physics \& NANOlab Center of Excellence, University of Antwerp,
 Groenenborgerlaan 171, B-2020 Antwerp, Belgium
}


\begin{abstract}
\noindent  
The superconducting properties of two recently synthesized ternary silicides with unit formula $X_4$CoSi ($X=$ Nb, Ta) are investigated through \textit{ab initio} calculations combined with Eliashberg theory. Interestingly, their crystal structure comprises interlocking honeycomb networks of Nb/Ta atoms. Nb$_4$CoSi is found to harbor better conditions for phonon-mediated superconductivity, as it possesses a higher density of states at the Fermi level, fostering stronger electron-phonon coupling. The superconducting critical temperatures ($T_c$) follow the same trend, with Nb$_4$CoSi having a twice higher value than Ta$_4$CoSi. Furthermore, the calculated $T_c$ values (6.0 K vs.\ 3.2 K) agree excellently with the experimentally obtained ones, establishing superconductivity in this new materials class as mediated by the electron-phonon coupling. Furthermore, our calculations show that the superconducting properties of these compounds do not simply correlate with the parameters of their honeycomb networks, contrary to proposals raised in the literature. Rather, their complete fermiology and phonon spectrum should be taken into account in order to explain their respective superconducting properties.   
\end{abstract}

\maketitle

\section{Introduction}

Ternary silicide intermetallic compounds are comprised of silicon together with two different metallic elements. They were found to harbor highly diverse physical properties, including heavy-fermion superconductivity in CeRhSi$_3$ \cite{PhysRevLett.95.247004}, two-gap superconductivity in $X_2$Fe$_3$Si$_5$ ($X=$ Sc, Lu) and Sc$_5$Ir$_4$Si$_{10}$ \cite{PhysRevLett.100.157001,doi:10.1088/1468-6996/9/4/044206}, and large magnetoresistance in NdCo$_2$Si$_2$ \cite{ROYCHOWDHURY2018625}. 

Recently, a new member of this materials family, Ta$_4$CoSi, was synthesized for the first time \cite{PhysRevB.106.134501}. Based on resistivity and magnetic susceptibility measurements, it was found to be a superconductor with a critical temperature ($T_c$) of 2.45 K. Nb$_4M$Si ($M=$ Ni, Co, Fe) compounds with the same crystal structure, characterized by the centrosymmetric tetragonal space group \textit{P4/mcc} (No. 124), were also found to display superconductivity \cite{PhysRevB.84.224518}. In the case of Nb$_4$CoSi, $T_c$ reaches 6.0 K, while the $T_c$'s of Nb$_4$FeSi (6.8 K) and Nb$_4$NiSi (7.7 K) are slightly higher. 

Interestingly, these $X_4M$Si bulk materials host honeycomb networks composed of niobium or tantalum atoms \cite{PhysRevB.106.134501,PhysRevB.84.224518}, though they are not occurring as distinct layers unlike the honeycomb boron networks in the two-gap superconductor MgB$_2$ \cite{Nagamatsu2001}. Other superconductors hosting honeycomb networks are mainly situated in low-dimensional systems, such as lithium- and calcium-doped graphene \cite{Profeta2012,doi:10.1073/pnas.1510435112,Chapman2016}, and gallenene \cite{Petrov_2021}. 

In their recent work on Ta$_4$CoSi, Zeng \textit{et al.} propose a linearly decreasing relation between $T_c$ and the interatomic distance within the honeycomb network, based on available values for a few superconducting compounds with a honeycomb network, ternary silicides and others \cite{PhysRevB.106.134501}. However, this apparent relation is rather counterintuitive, as biaxial tensile strain -- whereby the interatomic distances increase -- tends to have the opposite effect. Namely, the resulting attenuation of the interatomic electron clouds leads to phonon softening, which can enhance the electron-phonon coupling and $T_c$, according to Eliashberg theory. Systems for which this effect has been demonstrated include materials with a honeycomb network, such as doped graphene \cite{PhysRevLett.111.196802}, and monolayer MgB$_2$ in pristine and hydrogenated form \cite{PhysRevB.96.094510,PhysRevLett.123.077001}.

Here, using first-principles calculations of electronic structure, phonons, and their interaction, we demonstrate that superconductivity in this class of ternary silicides is conventional in nature. To this end, we show that the $T_c$ is correctly predicted by Eliashberg theory, indicating the absence of unconventional effects, like heavy-fermion superconductivity. We also set out to explain the difference in $T_c$ between Ta$_4$CoSi and Nb$_4$CoSi, and to elucidate the role of their respective honeycomb networks, and of the spin-orbit coupling, being stronger for tantalum than for niobium.

\section{Methods}

The calculations were performed within density functional theory (DFT), as implemented within the ABINIT code \cite{Gonze2020}, using the Perdew–Burke–Ernzerhof (PBE) functional \cite{PBE}. Fully relativistic optimized norm-conserving pseudopotentials from the PseudoDojo project were used to include spin-orbit coupling (SOC). The valence electron configurations of the used pseudopotentials are Nb-$4s^2 4p^6 4d^4 5s^1$, Ta-$5s^2 5p^6 5d^3 6s^2$, Co-$3s^2 3p^6 3d^7 4s^2$ and Si-$3s^2 3p^2$. For all the calculations, an energy cutoff value of 50 Ha for the plane-wave basis, and a $12 \times 12 \times 16$ $k$-grid were used. Gaussian smearing was used for the electronic occupations around the Fermi level ($E_F$), with a smearing width of 0.01 Ha. The crystal structures were relaxed so all force components were below $5 \times 10^{-6}$ Ha/bohr for each atom.  

To calculate the phonon dispersions and the electron-phonon coupling (EPC), we used density functional perturbation theory (DFPT) as implemented in ABINIT, using a $12 \times 12 \times 16$ electronic $k$-grid and a $3 \times 3 \times 4$ phononic $q$-grid. After thoroughly checking that its effect on the electronic structure near the Fermi level is very limited, SOC was omitted in the DFPT calculations, due to computational restrictions posed by the size of the unit cell. To characterize the superconducting state we employed the Migdal-Eliashberg theory for phonon-mediated superconductivity \cite{Eliashberg1, Eliashberg2, Eliashberg3}. Using the density of states at the Fermi level ($ N_F$), the matrix elements of the electron-phonon coupling ($g_{\textbf{k},\textbf{k}+\textbf{q}}^{\nu}$), and the phononic ($\omega_{\textbf{q}}^{\nu}$) and electronic ($\epsilon_{\textbf{k}}$) dispersions -- all of which were obtained from our \textit{ab initio} calculations -- we have evaluated the isotropic Eliashberg functions,
\begin{equation}
\alpha^2F(\omega)=\frac{1}{N_F}\sum_{\nu,\textbf{k},\textbf{q}}\left|g_{\textbf{k},\textbf{k}+\textbf{q}}^{\nu}\right|^2 \delta\left( \omega-\omega_{\textbf{q}}^{\nu}\right)\delta\left(\epsilon_{\textbf{k}} \right)\delta\left(\epsilon_{\textbf{k}+\textbf{q}} \right) ,
\label{eq:eliashberg_function}
\end{equation}
and the corresponding electron-phonon coupling (EPC) constants by 
\begin{equation}
\lambda=2\int_0^\infty \alpha^2F(\omega)\omega^{-1}d\omega .
\label{eq:lambda}
\end{equation}
The summations in Eq.\ \eqref{eq:eliashberg_function} were evaluated using gaussian integration with a smearing value of $10^{-4}$ Ha. We subsequently calculated the superconducting $T_c$ using the Allen–Dynes formula \cite{PhysRev.167.331, PhysRevB.12.905, ALLEN19831}. Here, the screened Coulomb interaction between Cooper-pair electrons was treated through the Morel-Anderson pseudopotential $\mu^*$ \cite{PhysRev.125.1263}, where a value of 0.13 was adopted, suitable for transition metal compounds with $d$-electrons \cite{Grimvall}.

\section{Crystal structure}

\begin{figure}[t]
                \includegraphics[width=\linewidth]{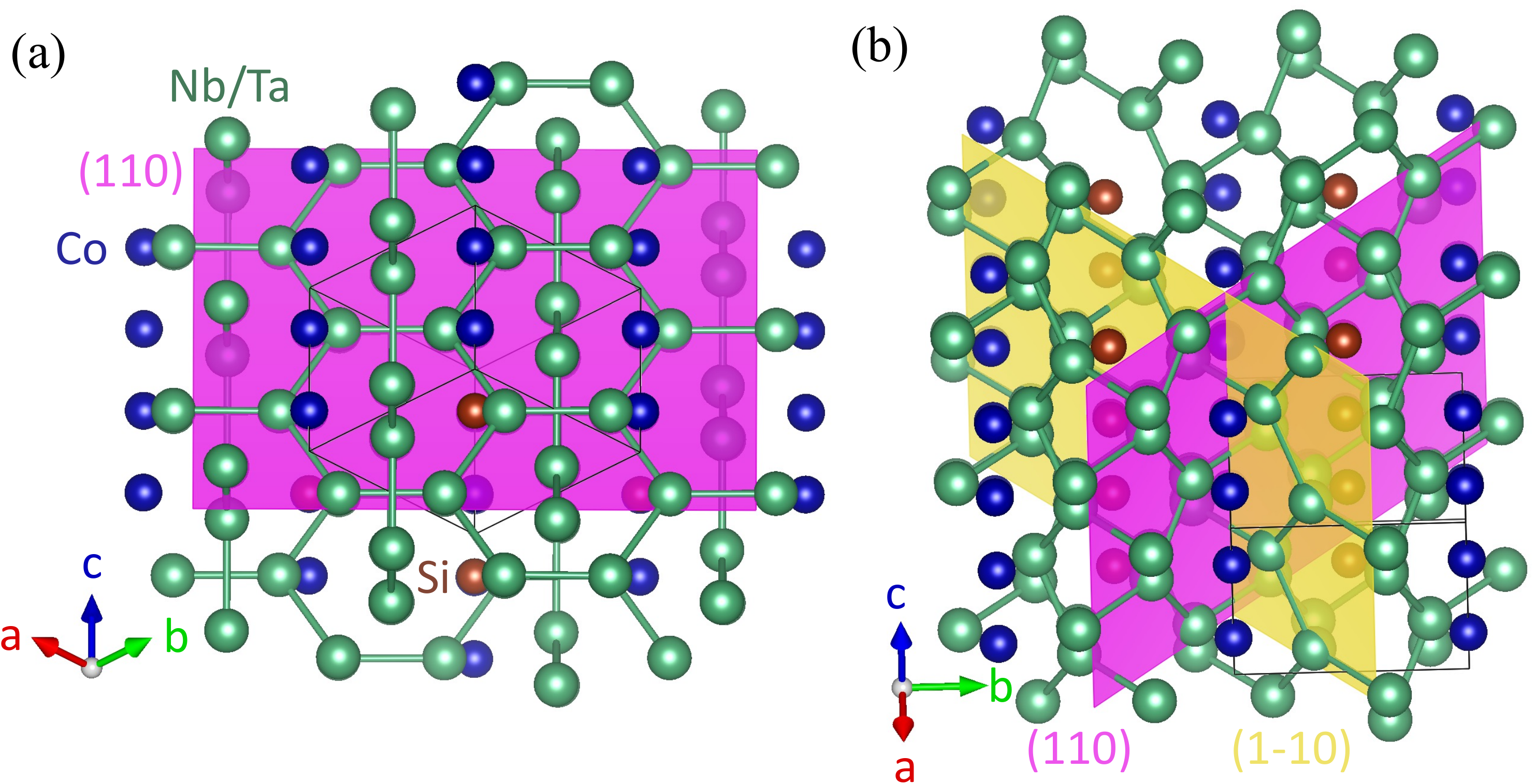}
                \caption{Crystal structure of Nb$_4$CoSi and Ta$_4$CoSi. (a) Honeycomb network of Nb/Ta atoms within the $(110)$ planes. (b) The two perpendicular, interlocking honeycomb networks within the $(110)$ and $(1\bar{1}0)$ planes.}
                \label{fig:structure}
\end{figure}

The crystal structure of the ternary silicides is displayed in Fig.\ \ref{fig:structure}. The unit cell consists of two $X_4$CoSi formula units, hence it contains twelve atoms. As mentioned in the introduction, these structures harbor networks of Nb/Ta atoms arranged in a honeycomb lattice. The first honeycomb network is located within the $(110)$ plane, as depicted in Fig.\ \ref{fig:structure}(a). As shown in Fig.\ \ref{fig:structure}(b), there is a second Nb/Ta honeycomb network in the $(1\bar{1}0)$ plane. As such, it is perpendicular to the first one, as well as interlocking with it, as can be clearly seen in Fig.\ \ref{fig:structure}(b). 

The calculated lattice parameters of the two investigated compounds are summarized in Table \ref{table1}, and compared to the experimental values available in the literature \cite{PhysRevB.84.224518,PhysRevB.106.134501}. An overall excellent agreement between theoretical and experimental values is obtained. Furthermore, we found that the inclusion of SOC does not affect the lattice parameters ($a$, $c$, $d_1$ and $d_2$) within a margin of 0.01 \AA, for neither Nb$_4$CoSi or Ta$_4$CoSi.

Since the Nb/Ta honeycomb networks are irregular, the hexagons possess two inequivalent sides, with lengths $d_1$ and $d_2$, as shown in Table \ref{table1}. Note that between Nb$_4$CoSi and Ta$_4$CoSi, $d_1$  increases slightly, while $d_2$ decreases -- a trend agreed upon by the calculations and experiments. As a result, there is a small change in the anisotropy parameter $d_1/d_2$, from 94.6\% in Nb$_4$CoSi to 95.6\% in Ta$_4$CoSi. So, rather than a uniform change in lattice parameters of the honeycomb network -- proposed in Ref.\ \citenum{PhysRevB.106.134501} to drive differences in $T_c$ -- there is a reduction of the anisotropy, going from Nb$_4$CoSi to Ta$_4$CoSi.

\begin{table}[b]
\centering
\begin{tabular}{lccccc}
\hline
Compound & Ref. & $a$ (\AA) & $c$ (\AA) & $d_1$ (\AA) & $d_2$ (\AA) \\\hline\hline
Nb$_4$CoSi & TW & 6.21 & 5.02 & 2.81 & 2.97 \\
Nb$_4$CoSi & EXPI & 6.16 & 5.06 & 2.78 & 2.99 \\\hline
Ta$_4$CoSi & TW & 6.21 & 4.99 & 2.82 & 2.95 \\
Ta$_4$CoSi & EXPII & 6.18 & 4.98 & 2.80 & 2.94  \\\hline
\end{tabular}
\caption{Comparison of calculated and experimental lattice parameters of Nb$_4$CoSi and Ta$_4$CoSi. Theoretical values from this work are indicated by `TW'. Experimental values stem from Ref.\ \citenum{PhysRevB.84.224518} (`EXPI') for Nb$_4$CoSi, and from Ref.\ \citenum{PhysRevB.106.134501} (`EXPII') for Ta$_4$CoSi. $d_1$ and $d_2$ are the $X-X$ ($X=$ Nb, Ta) distances within the honeycomb network.}
\label{table1}
\end{table}

\section{Electronic structure}

\begin{figure}[t]
                \includegraphics[width=\linewidth]{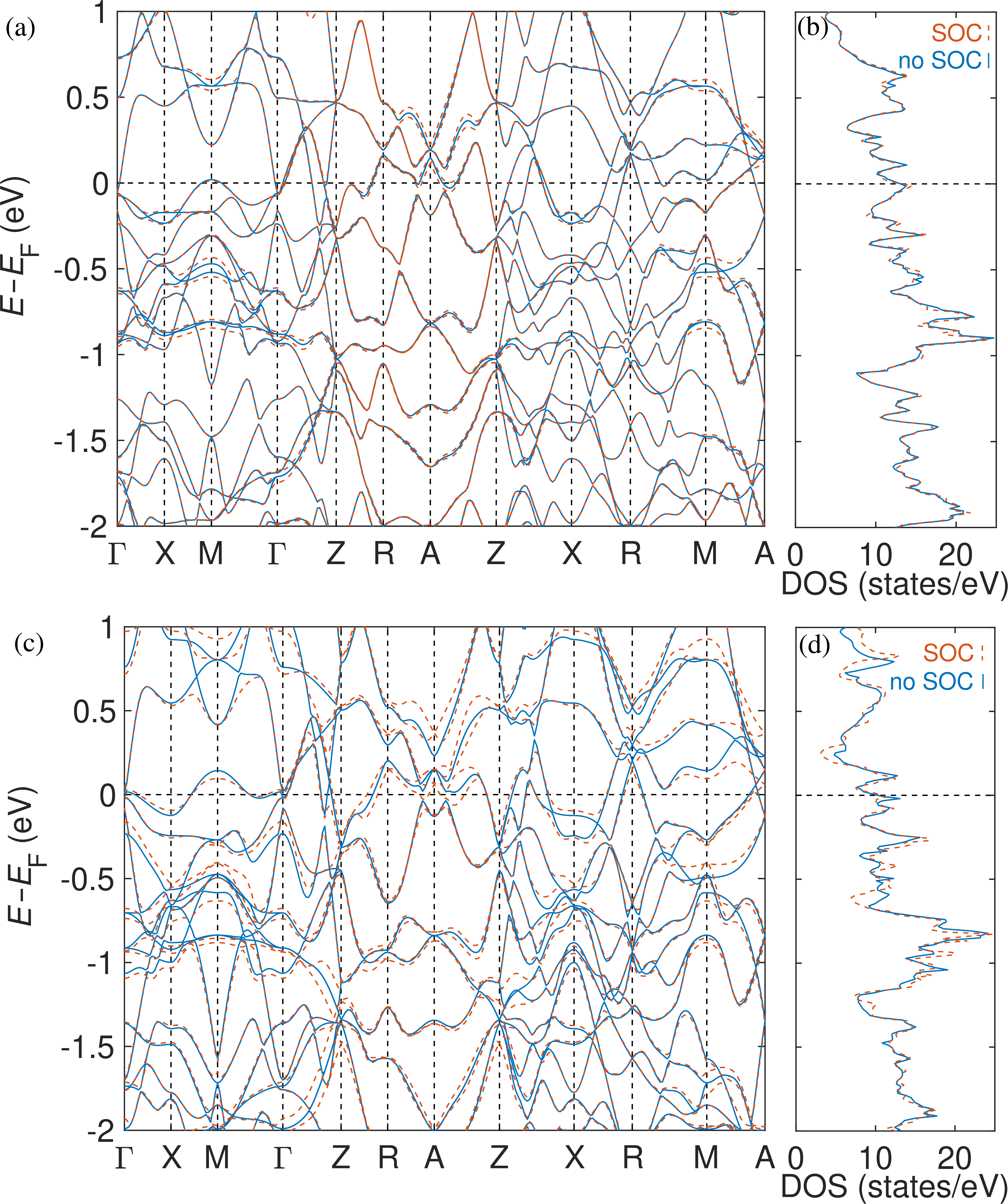}
                \caption{Electronic band structure of (a) Nb$_4$CoSi and (c) Ta$_4$CoSi. (b) The corresponding DOS of Nb$_4$CoSi and (d) Ta$_4$CoSi. Results are presented with and without inclusion of SOC, depicted by dashed orange lines and solid blue lines, respectively.}
                \label{fig:BS_DOS}
\end{figure}

\begin{figure*}[t]
                \includegraphics[width=\linewidth]{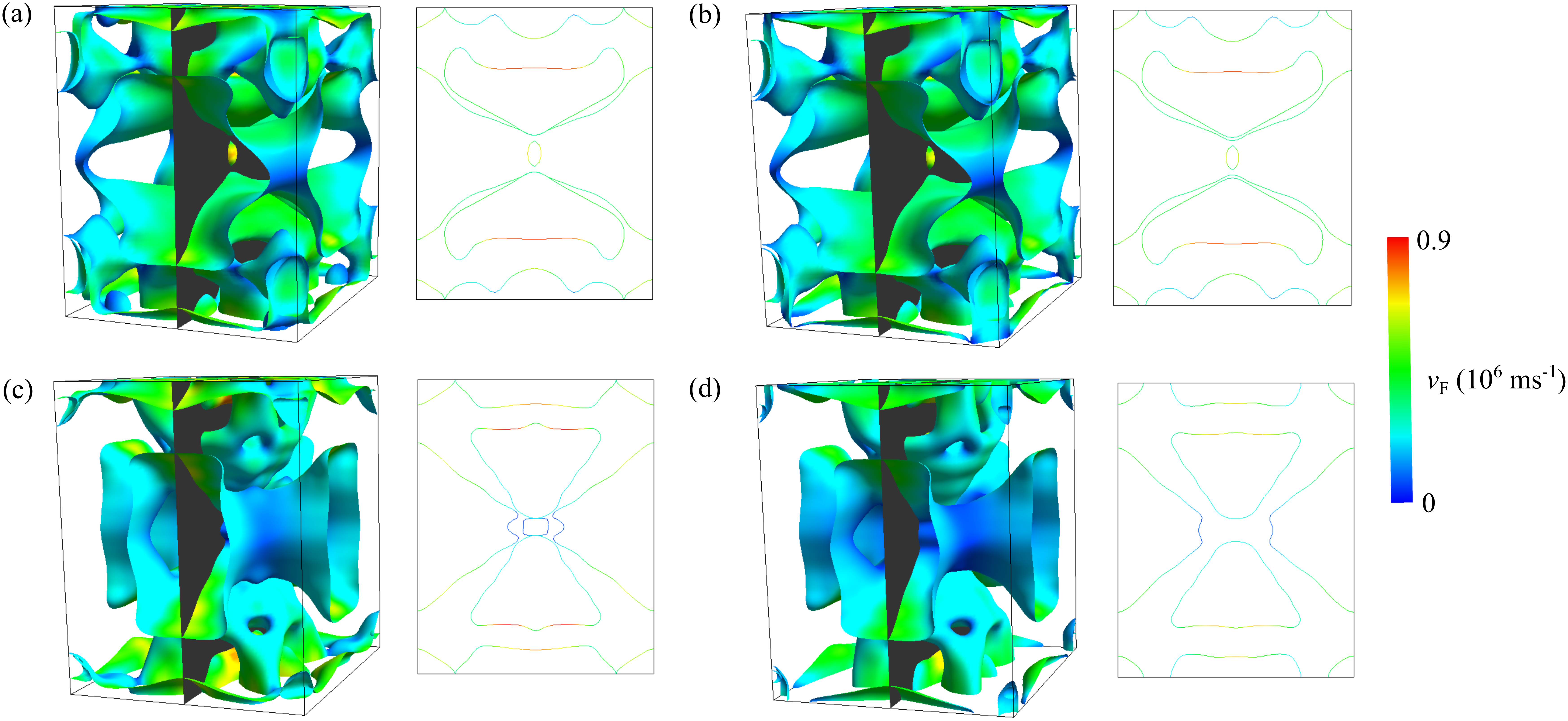}
                \caption{Fermi surface and its $(100)$ section (indicated by the grey plane), depicted within the $\Gamma$-centered Brillouin zone, for (a) Nb$_4$CoSi calculated without inclusion of SOC, and (b) with inclusion of SOC. Idem for (c) Ta$_4$CoSi calculated without inclusion of SOC, and (d) with inclusion of SOC. The colors indicate the Fermi velocities ($v_F$).}
                \label{fig:FS}
\end{figure*}

As the second step towards the superconducting properties of Nb$_4$CoSi or Ta$_4$CoSi, we now take a look at their electronic structure. The band structures of both compounds are shown in Fig.\ \ref{fig:BS_DOS}, calculated with and without inclusion of SOC. In the case of Nb$_4$CoSi (see Fig.\ \ref{fig:BS_DOS}(a)), the changes in the band structure due to SOC are very limited. As a result, the DOS profiles with and without SOC, shown in Fig.\ \ref{fig:BS_DOS}(b), are nearly identical. 

In the case of Ta$_4$CoSi, depicted in Fig.\ \ref{fig:BS_DOS}(c), more changes in the electronic structure occur upon inclusion of SOC -- due to the heavier atomic mass of Ta (with atomic number $Z=73$). However, these are mainly situated away from $E_F$ -- in both the valence and the conduction region. Near $E_F$, the effect of SOC is limited to some changes in the band effective masses, and a few minor band splittings (e.g., along the path Z--R). As a result, the DOS at $E_F$ is almost identical with and without SOC (see Fig.\ \ref{fig:BS_DOS}(d)). This enables us to evaluate the phonons, electron-phonon coupling, and the resulting superconducting properties without including SOC. Finally, we found that the DOS at $E_F$ is significantly higher in the case of Nb$_4$CoSi than for Ta$_4$CoSi, namely 13.10 states/eV compared to 10.22 states/eV. 

The corresponding Fermi surfaces are shown in Fig.\ \ref{fig:FS}, together with the Fermi velocities. Comparing the changes in the Fermi surface due to SOC, one again clearly observes only minor variations in both the sheet composition and Fermi velocities in the case of Nb$_4$CoSi (Fig.\ \ref{fig:FS}(a) vs.\ (b)). For Ta$_4$CoSi, more significant changes are found (Fig.\ \ref{fig:FS}(c) vs.\ (d)), notably around the Brillouin zone vertex (point A), and center ($\Gamma$).

\section{Phonons and electron-phonon coupling}

\begin{figure}[t]
                \includegraphics[width=\linewidth]{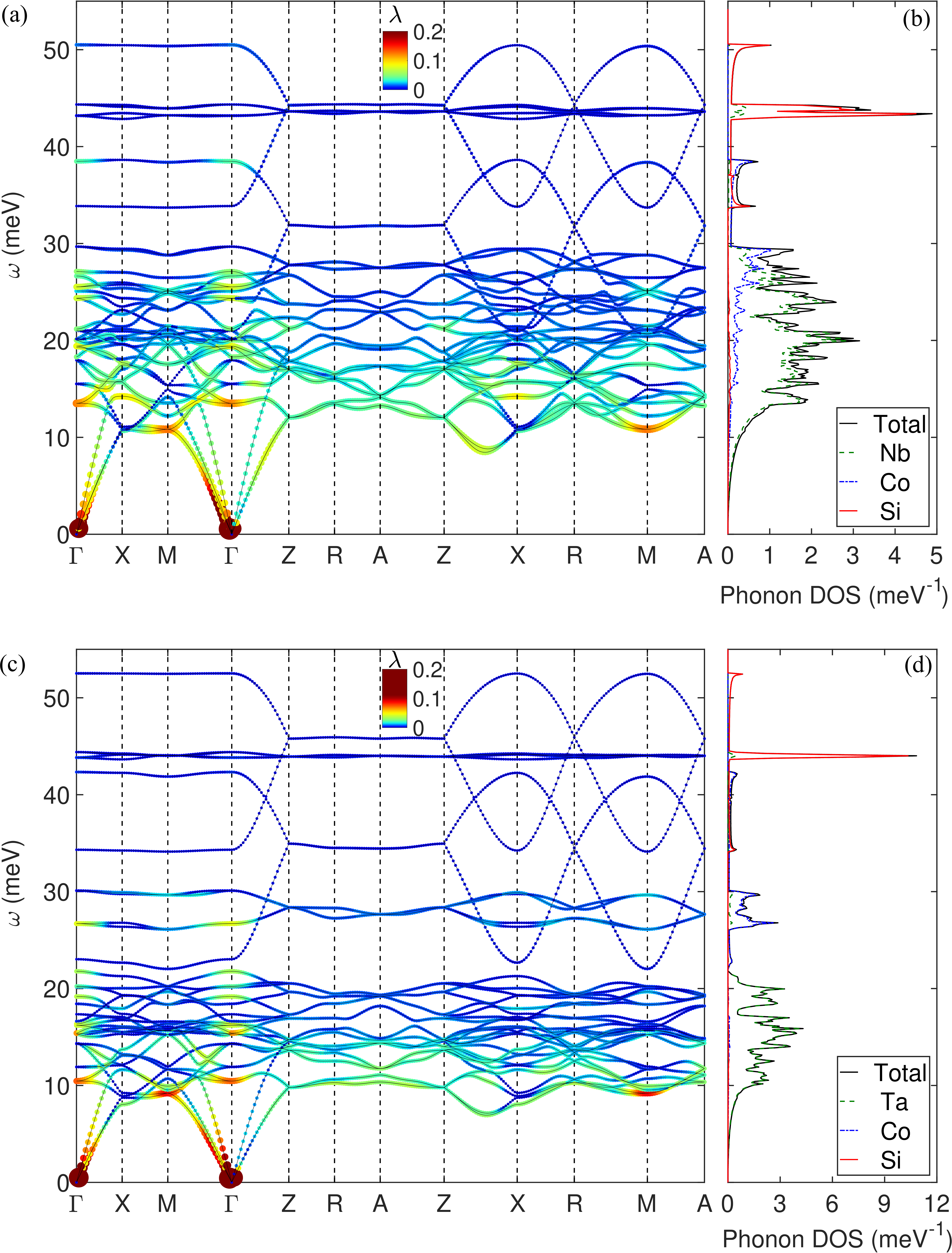}
                \caption{Phonon band structure of (a) Nb$_4$CoSi and (c) Ta$_4$CoSi. The phonon-branch ($\nu$) and \textit{q}-point-resolved electron-phonon coupling $\lambda_{q\nu}$ is indicated by colors, as well as by point sizes (proportional to $\lambda$). (b) The corresponding phonon DOS -- total as well as atom-resolved -- of Nb$_4$CoSi and (d) Ta$_4$CoSi.}
                \label{fig:PhBS_PhDOS}
\end{figure}

The vibrational properties of ternary silicides are interesting because of the presence of a very wide range of atomic masses, from very light elements like Si ($Z=14$), over intermediate mass elements like Co ($Z=27$), to heavy transition metals like Nb ($Z=41$) and Ta ($Z=73$). The phonon dispersions of Nb$_4$CoSi and Ta$_4$CoSi are depicted in Fig.\ \ref{fig:PhBS_PhDOS}(a) and (c) respectively, and the corresponding atom-resolved phonon DOS in (b) and (d). 

First of all, the phonon dispersions clearly demonstrate the dynamical stability of both compounds, without any phonon softening or Kohn anomalies that are known to occur in other transition metal compounds such as transition metal dichalcogenides (TMDs). Due to the differences in atomic masses of Nb and Ta, the phonon DOS's show different hybridization behavior. Namely Co hybridizes with both Nb and Si in Nb$_4$CoSi, whereas in Ta$_4$CoSi it only hybridizes with Si (to a very minor extent), leaving the Ta-based phonon modes entirely separated. Furthermore, several of the Si-based phonon modes have a remarkably flat dispersion, leading to pronounced peaks in the phonon DOS at higher energies in both compounds. 

The mode- and momentum-resolved EPC is depicted on top of the phonon dispersion in \ref{fig:PhBS_PhDOS}(a) and (c). It reaches its highest values for the lower Nb- and Ta-based modes. On the other hand, the Si-based modes host almost no EPC. Earlier analysis of the electronic structure has also revealed that the electronic states near $E_F$ have predominant transition metal $d$ character \cite{PhysRevB.84.224518,PhysRevB.106.134501}, hence the EPC in these ternary silicides is mainly driven by Nb and Ta.

\section{Superconductivity}

\begin{figure}[b]
                \includegraphics[width=\linewidth]{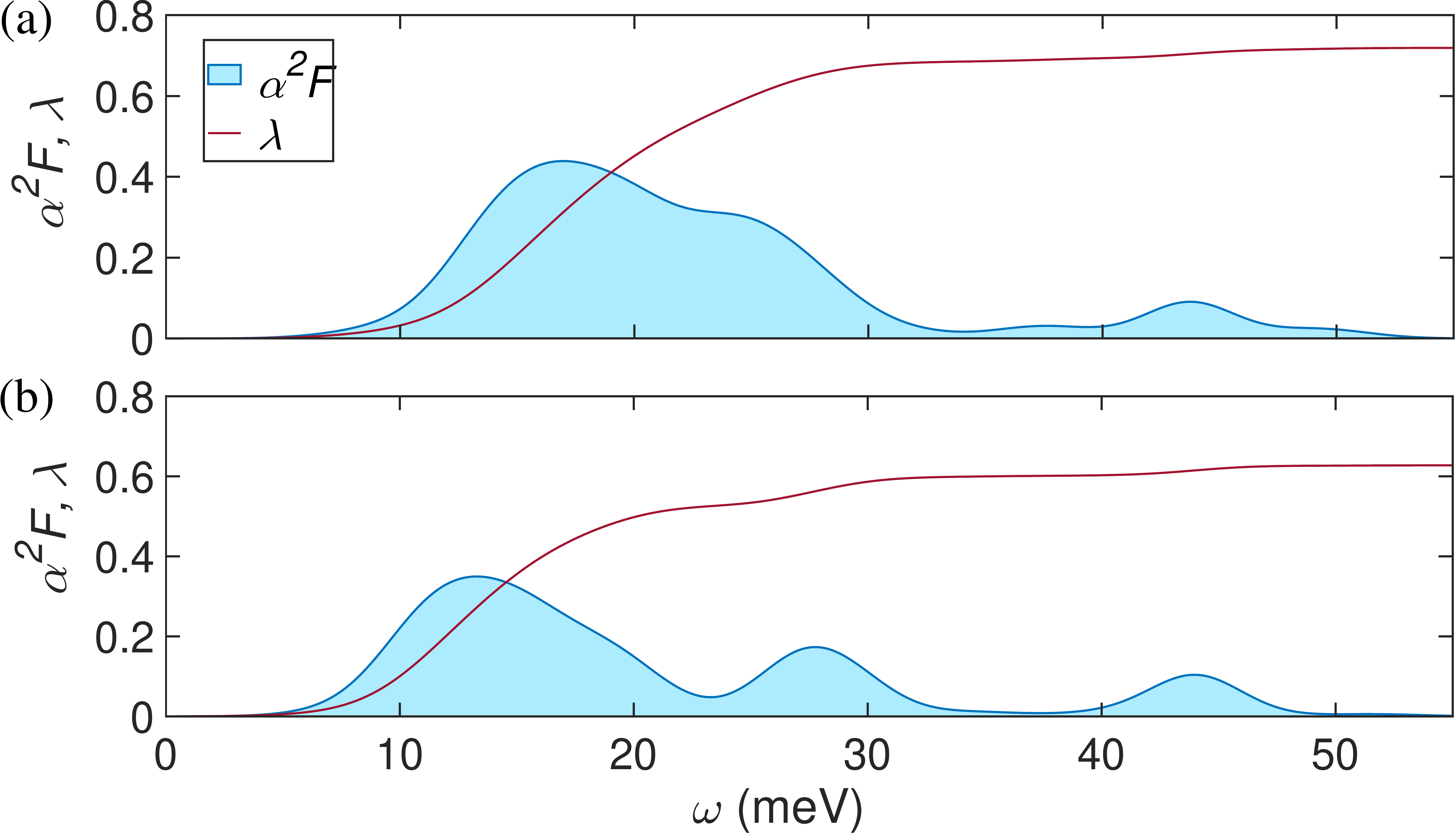}
                \caption{Eliashberg function ($\alpha^2F$) and electron-phonon coupling function ($\lambda$) of (a) Nb$_4$CoSi and (b) Ta$_4$CoSi.}
                \label{fig:eliashberg}
\end{figure}

\begin{table}[h]
\centering
\begin{tabular}{lccccc}
\hline
Compound & Ref. & $N_F$ (eV$^{-1}$) & $\omega_{log}$ (K) & $\lambda$ & $T_c$ (K) \\\hline\hline
Nb$_4$CoSi & TW & 13.10 & 209 & 0.72 & 6.0 \\
Nb$_4$CoSi & EXPI & - & - & - & 6.0 \\\hline
Ta$_4$CoSi & TW & 10.22 & 170 & 0.63 & 3.2 \\
Ta$_4$CoSi & EXPII & - & - & - & 2.45 \\ \hline
\end{tabular}
\caption{Superconducting properties of Nb$_4$CoSi and Ta$_4$CoSi. Theoretical values from this work are indicated by `TW'. Experimental values stem from Ref.\ \citenum{PhysRevB.84.224518} (`EXPI') for Nb$_4$CoSi, and from Ref.\ \citenum{PhysRevB.106.134501} (`EXPII') for Ta$_4$CoSi. The DOS at $E_F$ ($N_F$) is given per 12-atom unit cell. The Morel-Anderson pseudopotential used to calculate $T_c$ is $\mu^*=0.13$.}
\label{table2}
\end{table}

We now proceed to investigate the superconducting properties of Nb$_4$CoSi and Ta$_4$CoSi within Eliashberg theory. Previous analysis of their superconducting $T_c$'s was limited to models based on the Debye temperature ($\theta_D$) \cite{PhysRevB.106.134501}, whereby $T_c \propto \theta_D$. It should be noted that this model is not very adequate for materials with an extended unit cell, as it is limited to acoustic modes. Instead, the logarithmically averaged phonon frequency $\omega_{log}$, arising in Eliashberg theory, forms the core proportionality with $T_c$, as established in the Allen-Dynes adaptation of McMillan's formula \cite{PhysRev.167.331,PhysRevB.12.905,ALLEN19831}. 

As the first step, we calculated the Eliashberg spectral functions ($\alpha^2F$) from our DFPT calculations according to Eq.\ \eqref{eq:eliashberg_function}. The results, depicted in Fig.\ \ref{fig:eliashberg}, show that the main contributions to the Eliashberg functions occur below 30 meV in both ternary silicide compounds, and stem from the Nb/Ta- and Co-based phonon modes. For the case of Nb$_4$CoSi (Fig.\ \ref{fig:eliashberg}(a)), their contributions form a single dome, with a kink around 23 meV, reflecting the depletion in the phonon DOS at that energy (cf.\ Fig.\ \ref{fig:PhBS_PhDOS}(a)). The case of Ta$_4$CoSi (Fig.\ \ref{fig:eliashberg}(b)) presents a more pronounced separation of the contributions of Ta and Co, leading to two distinct domes in the Eliashberg function below 30 meV. This separation originates from the higher atomic mass of Ta, pushing its phonon modes to lower energies. This effects a reduction of the hybridization with the Co-based phonon modes (as explained in the preceding section). 

The corresponding electron-phonon coupling functions ($\lambda$), obtained from the Eliashberg functions using Eq.\ \eqref{eq:lambda}, are also shown in Fig.\ \ref{fig:eliashberg}. It is apparent that the total $\lambda$ of Nb$_4$CoSi (0.72) significantly exceeds that of Ta$_4$CoSi (0.63). This stronger EPC in Nb$_4$CoSi clearly correlates with its higher $N_F$. Together with its higher $\omega_{log}$, this results in a higher $T_c$, as obtained from the Allen-Dynes formula. The obtained $T_c$ of Nb$_4$CoSi is 6.0 K, compared to 3.2 K for Ta$_4$CoSi, in perfect agreement with their experimental values of 6.0 K \cite{PhysRevB.106.134501} and 2.45 K \cite{PhysRevB.106.134501}, respectively. These results are summarized in Table \ref{table2}.

\section{Conclusions}

We have performed an \textit{ab initio} investigation of the structural, electronic, vibrational and superconducting properties of two intermetallic ternary silicide compounds with unit formula $X_4$CoSi, differing only by the $X$ element, being Nb or Ta. These recently synthesized compounds have been attracted significant interest in view of their transition metal honeycomb networks and experimentally observed superconductivity \cite{PhysRevB.106.134501}.    

We found that both compounds harbor a rich fermiology in the vicinity of the Fermi level, characterized by a multitude of intersecting Fermi sheets (see Fig.\ \ref{fig:FS}). Between them, Nb$_4$CoSi has a 28\% higher DOS at the Fermi level, leading to enhanced EPC compared to Ta$_4$CoSi (14\% higher). Here, the higher mass of the Ta atoms mitigates the discrepancy in the EPC between the two compounds. Namely, it shifts the Eliashberg function to a lower phonon energy range (see Fig.\ \ref{fig:eliashberg}), where it contributes more effectively to the EPC constant ($\lambda$). Nevertheless, due to the exponential dependence of $T_c$ on $\lambda$, this leads to a marked difference in $T_c$ values, 6.0 K vs.\ 3.2 K for Nb$_4$CoSi and Ta$_4$CoSi, respectively, agreeing to a high degree with the experimentally obtained values (6.0 and 2.45 K \cite{PhysRevB.84.224518,PhysRevB.106.134501}). 

Due to the excellent agreement with the experimental $T_c$ values, our calculations provide strong evidence for the conventional nature of superconductivity in these compounds. We also found that the main factor driving the difference in $T_c$ between the Nb- and the Ta-based compound is their intrinsic difference in DOS at their respective Fermi levels. No direct link to the lattice parameters of their transition metal honeycomb networks was found, contrary to recent proposals raised in the literature \cite{PhysRevB.106.134501}. The reasons for the absence of such link are that (i) the honeycomb network is irregular and does not change uniformly between different silicide compounds (rather, its anisotropy changes) and (ii) our Eliashberg calculations show that the electron-phonon coupling results from a multitude of electronic and phononic states, not just those related to the honeycomb networks, indicating the inadequacy of a simple Debye-type model \cite{PhysRevB.106.134501}. We also found that the role of SOC in the normal-state electronic as well as the superconducting properties is limited for both compounds (and especially so in the case of Nb$_4$CoSi). 

These results emphasize the importance of taking into account the complete fermiology and phonon spectrum to calculate the superconducting properties of complex intermetallic compounds, as well as the significant role therein of the atomic masses of the constituent transition metal elements. 

\begin{acknowledgments}
\noindent J.B. is a senior postdoctoral fellow of Research Foundation-Flanders (FWO, fellowship No.\ 12ZZ323N). The computational resources and services were provided by the VSC (Flemish Supercomputer Center), funded by the FWO and the Flemish Government -- department EWI. 
\end{acknowledgments}

\bibliography{Refs}

\end{document}